\documentclass[12pt]{article}

\usepackage{amstext,amsmath,amssymb,amsfonts,bbm}
\usepackage[latin1]{inputenc}
\usepackage{epsfig}
\usepackage{hyperref}
\usepackage{amsthm}
\usepackage{subfigure}
\usepackage{color}
\usepackage{multirow}
\newcommand{\bea}{\begin{eqnarray}}	
\newcommand{\eea}{\end{eqnarray}}
\newcommand{\cG}{{\cal G}}
\newcommand{\cV}{{\cal V}}
\newcommand{\cC}{{\cal C}}

\newcommand{\cB}{{\cal B}}

\newcommand{\mB}{\mathfrak{B}}
\newcommand{\ZZ}{\mathbb{Z}}

\newtheorem{lemma}{Lemma}

\newtheorem{definition}{Definition}

\begin{document}

\title{Colored Group Field Theory}

\author{Razvan Gurau\footnote{Perimeter Institute for Theoretical Physics Waterloo, ON, N2L 2Y5, Canada } }

\maketitle

\begin{abstract}
\noindent
Group field theories are higher dimensional generalizations of matrix models. Their Feynman graphs are fat and in addition to vertices, edges and faces, they also contain higher dimensional cells, called bubbles.

In this paper, we propose a new, fermionic Group Field Theory, posesing a color symmetry, and take the first steps in a systematic study of the topological properties of its graphs. Unlike its bosonic counterpart, the bubbles of the Feynman graphs of this theory are well defined and readily identified. We prove that this graphs are combinatorial cellular complexes. We define and study the cellular homology of this graphs. Furthermore we define a homotopy transformation appropriate to this graphs. Finally, the amplitude of the Feynman graphs is shown to be related to the fundamental group of the cellular complex.

\end{abstract}

\section{Introduction: Group Field Theory}

Group field theories (GFT) \cite{boulatov,laurentgft,iogft,iogft2,Magnen:2009at} (that is quantum field theories over group manifolds) are generalizations of matrix and tensor models \cite{mm,gross,ambjorn,Sasakura:1990fs}). They arise naturally in several approaches to quantum gravity, like the discrete approaches (Regge calculus \cite{williams}, dynamical triangulations \cite{DT}) or the spin foam models \cite{SF} (see also \cite{libro} for further details).

As a quantum field theory, a GFT is defined by an action functional depending on a field. For some group $G$, the field in a $n$ dimensional GFT is a scalar $\phi:G^{\otimes n-1}\rightarrow \mathbb{C}$, invariant under the left group action.

The action of a GFT writes (see \cite{DP-P})
\bea \label{eq:action}
&&S=\frac{1}{2}\int [dg]
\bar \phi_{\alpha_0,\dots, \alpha_{n-1}}
K(g_{\alpha_0},\dots, g_{\alpha_{n-1}};g_{\alpha'_0},\dots g_{\alpha'_{n-1}} )
\phi_{\alpha'_0,\dots, \alpha'_{n-1}} 
\nonumber\\
&&+
\frac{\lambda}{n+1}\int [dg] V(g_{\alpha^{0}_0}, \dots, g_{\alpha^{n+1}_{n-1}})
\phi_{\alpha^{0}_0, \dots, \alpha^{0}_{n-1}} \dots \phi_{\alpha^{n+1}_0, \dots,\alpha^{n+1}_{n-1}}\;,
\eea
where we used the shorthand notations $\phi(g_{\alpha_0},\dots g_{\alpha_{n-1}})=\phi_{\alpha_0,\dots, \alpha_{n-1}}$, and $\int [dg]$ for the integral over the group manifold with Haar measure of all group elements appearing in the arguments of the integrand.

In $n$ dimmensions, the vertex operator $V$ encodes the connectivity dual to a $n$ dimensional simplex, whereas the propagator $K$ encodes the connectivity dual to the gluing of two $n$ dimensional simplexes along a $n-1$ dimensional subsimplex. Thus the field $\phi$ is associated to the $n-1$ dimensional simplexes, and the group elements $g$ to the $n-2$ dimensional simplexes.

The Feynman graphs of a GFT are dual to cellular complexes. The cells of this complexes are $n$D\footnote{Thruout this paper we will use the notation $n$D to signify ``$n$ dimensional''.} simplexes. They are glued together (as dictated by the propagator K) along $n-1$D simplexes of their boundary. These complexes will henceforth be referred to as ``space-times complexes''. Consequently, GFT is a combinatorial background independent theory which generates possible space-times backgrounds as duals to its Feynman diagrams. The topological properties of these space-time complexes, mainly in the spin foam formulation, has been abundantly studied, and for manifold space-times new topological invariants \cite{Turaev:1992hq} have been defined.

The rationale behind associating group elements to $n-2$D simplexes is that the Feynman amplitude of a give graph of this simple combinatorial model turns out to equal the partition function of a BF theory discretized on the space-time complex. As BF theory becomes Einstein gravity after implementing the Plebanski constraints, it is natural to suppose that choosing more complex vertex and propagator operators one will recover the partition function of discretized gravity.

In fact, in most of the particular GFT models explored so far \cite{barrett,EPR,Etera, FK} weights have been associated to the operators $K$ and $V$, on top of the mere combinatorial data, according to some procedure to implement this constraits. The semiclassical limit of such models has been investigated with encouraging results \cite{semicl,gravit} .

In this paper we take a different approach to GFT. As are interested in the combinatorial and topological aspects of the Feynman diagrams of the action (\ref{eq:action}), we will chose the purely combinatoric operators $K$ and $V$ which lead to the partition function of BF theory.

Furthermore as we take a quantum field theory approach to GFT's, we do not insist on the topological properties of the space-time complexes, but on those of the graphs themselves!

Our study is motivated by the following analogy with matrix models. Even though only identically distributed matrix models are topological in some scaling limit (\cite{mm}), it turns out that the power counting of more involved models (like for instance the Gross-Wulkenhaar model \cite{GW,GW1}) is again governed by topological data. In fact, from a quantum field theory perspective one can argue that the only matrix models for which some renormalization procedure can be defined are those with a topological power counting! Furthermore, the non trivial fixed point of the Gross-Wulkenhaar model \cite{razvan,Geloun:2008zk} opens up the intriguing possibility that GFTs are UV complete quantum field theories!

The natural first step in addressing more realistic GFT is to first clarify the question of the topology and combinatorics of simple topological models. In this paper we will deal with the simplest (as far as the topology and combinatorics of the Feynman diagrams are concerned) GFT model, defined in section \ref{sec:newmodel}. This new, fermionic, GFT has a surprising $SU(n+1)$ color symmetry\footnote{A very non natural colored bosonic model can be constructed which does not posses this symmetry.}. We then undertake the first steps in a systematic study of the topological properties of its graphs. The latter are detailed in in section \ref{sec:homology} where a cellular complex structure {\it of the graph itself} is defined. The ``bubble'' homology of this cellular complex, defined by a boundary operator related to the one of \cite{FreiGurOriti}, is detailed in section \ref{sec:homology}. The amplitudes of the graphs are subsequently related to the first homotopy group of the graph complex in section \ref{sec:homotopy}, and a necessary condition for a graph to be homotopically trivial is derived. Section \ref{sec:conclusion} draws the conclusion of our work and section \ref{sec:app} presents explicit computations of homology groups for several examples of graphs.

\section{The Fermionic Colored GFT}\label{sec:newmodel}

Consider, in stead of an unique bosonic field, $n+1$ {\bf Grassmann} fields, $\psi^0,\dots,\psi^n:G^{\otimes n-1}\rightarrow \mathbb{C}$ 
\bea
\{\psi^i,\psi^j\}=0 \; ,
\eea
with hermitian conjugation 
\bea
\psi\rightarrow \bar{\psi} \text{ such that } \overline{\psi^1\psi^2}=-\bar{\psi^2}\bar{\psi^1},
\quad \bar{\bar{\psi}}=-\psi \; .
\eea
The fields $\psi$ are chosen with no symmetry properties under permutations of the arguments, but are all invariant under simultaneous left action of the group on all there arguments. The upper index $p$ denotes the color of the field $\psi^p$.

A (global) color transformation is an internal rotation $U\in SU(n+1)$ on the grassmann fields
\bea
(\psi^i)'=U^{ij} \psi^j,\quad (\bar{\psi^i})'=\bar{\psi^j} (U^{ij})^*=\bar{\psi^j} (U^{-1})^{ji}
\; .
\eea
The only quadratic form invariant under color transformation is
\bea
\sum_p\bar{\psi}^p \psi^p \; .
\eea

The interaction is a monomial in the fields. The only monomial in $\psi$ invariant under color rotation is
\bea
 \psi^0 \dots \psi^n \; ,
\eea
as
\bea
 (\psi^0)'\dots (\psi^n)'=U^{0i_0}\dots U^{ni_n}\psi^{i_0}\dots\psi^{i_n}=\det(U) \psi^0 
\dots \psi^n \;.
\eea

The hermitian GFT action of minimal degree, invariant under (global) color rotation is

\bea\label{eq:admis}
S= \sum_p\int [dg] \; \bar{\psi^p} \psi^p 
+\int [dg] \; \psi^{0} \psi^{1}\dots \psi^{p} 
+\int [dg] \; \bar{\psi^{0}}\bar{\psi^{1}} \dots \bar{\psi^{p}} \; ,
\eea
where the arguments of $\psi^p$ and $\bar{\psi^{p}}$ in the interaction terms are chosen to reproduce the combinatorics of the GFT vertex (see \cite{DP-P}), that is, denoting $g_{pq}=g_{qp}$ the group associated to the strand connecting the halflines $p$ and $q$,
\bea
\psi^{p}(g_{p-1p},g_{p-2p},\dots, g_{0p},g_{pn}, \dots g_{pp+1}) \; .
\eea

\begin{figure}[htb]
\centering{
\includegraphics[width=60mm]{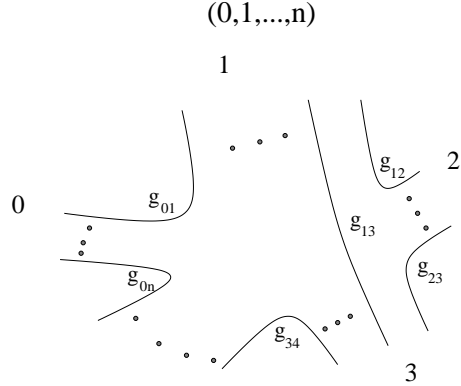}}
\caption{nD vertex.}
\label{fig:nDvertex}
\end{figure}

The intercation part of the action of equation (\ref{eq:admis}) has two terms. We call the vertex involving only $\psi$'s the positive vertex and represent it like in figure \ref{fig:nDvertex}. The second term, involving only $\bar{\psi}$ is similar, but the colors turn anticlockwise around it. The vertices have a detailed internal structure encoding the arguments $g$ of the fields and there connectivity. Each of these $g$'s corespond to a strand in figure \ref{fig:nDvertex}).

\begin{figure}[htb]
\centering{
\includegraphics[width=80mm]{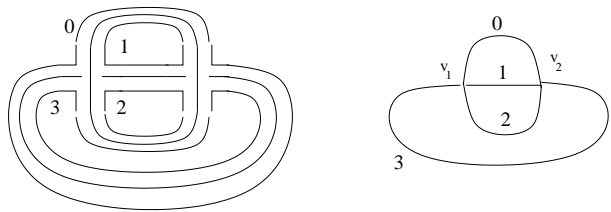}}
\caption{Stranded and colored graph.}
\label{fig:exemplu}
\end{figure}

From equation (\ref{eq:admis}) we conclude that the propagator of the model is formed of $n$ parallel strands and allways connects two halflines of the same color, one on a positive vertex and one on a negative vertex. As the structure of the vertex and propagator is fixed and no permutation of the arguments is allowed, we can encode all the relevant information of a graph in a colored graph with point vertices and colored lines. The strand structure is fixed once the colored graph is given. We orient all lines from positive to negative vertices. In figure \ref{fig:exemplu} we give an example of a 3 dimensional graph drawn either in all the detail or as colored graph with four colors.

A Feynman graph in which the strand structure is explicit will hencefroth be called a {\it stranded} graph, whereas the simplified graph in which only the colors are explicit will be called a {\it colored} graph.

\section{Bubble Homology in Colored GFT}\label{sec:homology}

The strand structure of the vertices and propagators render the Feynman graphs of our model topologically very rich. This should come as no surprise, as in three dimensions for instance, the space-times duals to graphs include not only all orientable piecewise linear three dimensional manifolds (see \cite{crystal} and references therein), but also pseudomanifolds. In this paper we take a new approach to study this topology: we will not focus on the topology of ``space-times'' dual to graphs, but we will concentrate solely on the topological structure of the graphs themselves!

For a graph, consider all {\it connected} subgraph made only of lines of certain, chosen, colors. We call such a graph a {\it bubble} and denote it $\cB^{\cC}_{\cV}$, where $\cC$ is the {\it ordered} set of colors of the lines in the bubbles and $\cV$ is the set of vertices. The graph itself is not considered a bubble. 

Thus for the example in figure \ref{fig:exemplu} we have the  subgraphs with three colors, $\cB^{012}_{v_1v_2}$, $\cB^{013}_{v_1v_2}$, $\cB^{023}_{v_1v_2}$ and $\cB^{123}_{v_1v_2}$, those with two colors $\cB^{01}_{v_1v_2}$, $\cB^{02}_{v_1v_2}$, $\cB^{03}_{v_1v_2}$, $\cB^{12}_{v_1v_2}$, $\cB^{13}_{v_1v_2}$, $\cB^{23}_{v_1v_2}$, those with one color $\cB^0_{v_1v_2}$, $\cB^1_{v_1v_2}$, $\cB^2_{v_1v_2}$, $\cB^3_{v_1v_2}$, and finally those with zero colors $\cB_{v_1}$, $\cB_{v_2}$. Note that the bubbles can also be drawn as graphs with strands. For instance in figure \ref{fig:exemplu}, the stranded graph of the bubble $\cB^{012}_{v_1v_2}$ is obtained by deleting all strands belonging to the line $3$. As such we see that the stranded graph of a bubble with $p$ colors (that is a $p$D bubble) is in fact itself GFT graph.

The subgraphs with zero colors are the vertices of the initial graph. The subgraphs with one color are the lines (as on any vertex we have only one halfline of a certain color). The subgraphs with two colors are cycles of lines along which the colors alternate. They are sometimes the ``faces'' of the graph. The GFT graphs with three colors are ribbon graphs and the reader can easily check that the faces defined here coincide with the usual notion of face for graphs of matrix models. The higher dimensional bubbles are the natural generalization of faces for higher dimmensions.

These bubbles are the building blocks of a combinatorial graph complex and generate an associated homology. Denote the set of bubbles with $p$ colors (that is $|\cC_{\cB}|=p$) by $\mathfrak{B}^{p}$, and define the $p$'th chain group as the finitely generated group $C_p(\cG)=\{\alpha_p\}$,
\bea
 \alpha_p=\sum_{{\cal B}^{\cC}_{\cV}\in \mathfrak{B}^p} c^{\cC}_{\cV}
 \cB^{\cC}_{\cV}\quad 
, c^{\cC}_{\cV} \in \mathbb{Z}  \; .
\eea
These chain groups define homology groups via a boundary operator defined as follows.

\begin{definition}\label{def:bound}
The $p$'th boundary operator $d_p$ acting on a $p$ bubble $\cB^{\cC}_{\cV}$ 
 with colors $\cC=\{i_{1},\dots i_{p}\}$ is
\begin{itemize} 
\item for $p\ge 2$,
\bea \label{eq:boundary}
d_p(\cB^{\cC}_{\cV})=
\sum_{q}(-)^{q+1} \sum_{ \stackrel{ {\cal B'}^{\cC'}_{\cV'}\in \mathfrak{B}^{p-1} } 
{ \cV'\subset \cV \; \cC' = \cC \setminus i_{q}} }  
{\cal B'}^{\cC'}_{\cV'} \;,
\eea
which associates to a $p$D bubble the alternating sum of all $p-1$D bubbles formed by subsets of its vertices. 

\item for $p=1$, as the lines $\cB^i_{v_1v_2}$ connect a positive vertex (say $v_1$) to a negative one, say $v_2$
\bea
 d_1 \cB^i_{v_1v_2}= \cB_{v_1}-\cB_{v_2} \;.
\eea
 \item for $p=0$, $d_0\cB_v=0$.
\end{itemize}
\end{definition}
These boundary operators give a well defined homology as
\begin{lemma}
The boundary operators respect $d_{p-1}\circ d_{p}=0$. 
\end{lemma} 
{\bf Proof:}
To check this consider the application of two consecutive boundary operators on a $p$ bubble
\bea
&&d_{p-1}d_p(\cB^{\cC}_{\cV}) = \sum_{q}(-)^{q+1} \sum_{ \stackrel{ {\cal B'}^{\cC'}_{\cV'}\in \mathfrak{B}^{p-1} } 
{ \cV'\subset \cV \; \cC' = \cC \setminus i_{q}} }  d_{p-1}\cB'^{\cC'}_{\cV'}
\\
&&=\sum_{q}(-)^{q+1} \sum_{ \stackrel{ {\cal B'}^{\cC'}_{\cV'}\in \mathfrak{B}^{p-1} } 
{ \cV'\subset \cV \; \cC' = \cC \setminus i_{q}} } 
\Big{[}
\sum_{r<q} (-)^{r+1}\sum_{ \stackrel{ {\cal B''}^{\cC''}_{\cV''}\in \mathfrak{B}^{p-2} } 
{ \cV''\subset \cV' \; \cC'' = \cC' \setminus i_{r}} } \cB''^{\cC''}_{\cV''} +\nonumber\\
&&\sum_{r>q} (-)^{r}\sum_{ \stackrel{ {\cal B''}^{\cC''}_{\cV''}\in \mathfrak{B}^{p-2} } 
{ \cV''\subset \cV' \; \cC'' = \cC' \setminus i_{r}} } \cB''^{\cC''}_{\cV''}
\Big{]} \;,
\eea
where the sign of the second term changes, as $i_{r}$ is the $r-1$'th color of
$\cC'\setminus i_q$ if $q<r$. The two terms cancel if we exchange $q$ and $r$ in the second term.

\qed

The operators $d_p$ provide direct access to the attaching maps of $p$ cells to $p-1$ cells, therefore
\begin{lemma}
 A colored GFT graph is an abstract cellular complex, with cells the $pD$ bubbles 
and attaching maps induced by the boundary operator of definition \ref{def:bound}.
\end{lemma}

Note that in general this combinatorial complex is not a CW complex. However it becomes one if all $p$D bubbles are homeomorphic with the boundary of a $p$ dimensional disk. In 3D this is the case if all 3 colored bubbles are planar, as such graphs are homeomorphic with the sphere $S^2=\partial D^3$ and the faces and lines are allways homeomorphic with $S^1=\partial D^2$ and 
$S^0=\partial D^1$ respectively.

\section{Minimal and Maximal Homology Groups}

Before proceding to analyze the Feynman amplitudes of colored GFT we will give some generic properties of the bubble homology induced by the boundary operator (\ref{eq:boundary}). Detailed examples of homology computations for graphs are presented in the appendix \ref{sec:app}.

First, by definition $d_0$ acting on zero bubbles is zero. Thus, for any graph,
\bea
\text{ker}(d_0)=\bigoplus_N\ZZ \;,
\eea
where $N=|\mathfrak{B}^0|$ is the number of vertices (0 bubbles) of the graph. Our first result concerns the minimal homology group of a colored graph.

\begin{lemma} \label{lem:d1}
For connected closed graphs $H_0=\ZZ$.
\end{lemma}
{\bf Proof}
The operators $d_1$ acting on a one chain is
\bea\label{eq:d1}
d_1\alpha_1=\sum_{\cB^{i}_{v_1v_2}\in \mathfrak{B}^1} c^{i}_{v_1v_2} d_1 \cB^{i}_{v_1v_2} \;.
\eea
The matrix of $d_1$ is then incidence matrix of the oriented graph, where a line enters its positive endvertex and exists from its negative end vertex
\bea
 \Gamma_{\cB_v \cB^i_{v_1v_2}}=\begin{cases}
             1, &\text{ if } \cB^i_{v_1v_2} \text{ enters } \cB_v \\
             -1,&\text{ if } \cB^i_{v_1v_2} \text{ exits } \cB_v \\
	     0, &\text{ else }
           \end{cases} \;,
\eea

We will compute the $\text{Im}(d_1)$ using a contraction procedure. We start by choosing a line $\cB^i_{v_1v_p}$ connecting the two distinct vertices $v_1$ and $v_p$. We collect all terms containing either $\cB_{v_1}$ or $\cB_{v_p}$ in equation (\ref{eq:d1}) to get
\bea\label{eq:terms}
d_1\alpha_1&=& \cB_{v_1} \Gamma_{\cB_{v_1}\cB^i_{v_1v_p}} c^{i}_{v_1v_p}+
\sum_{v_q\neq v_p} \cB_{v_1} \Gamma_{\cB_{v_1}\cB^i_{v_1v_q}} c^{i}_{v_1v_q} \nonumber\\
&+& \cB_{v_p} \Gamma_{\cB_{v_p}\cB^i_{v_1v_p}} c^{i}_{v_1v_p} +
\sum_{v_r\neq v_1} \cB_{v_p} \Gamma_{\cB_{v_p}\cB^i_{v_pv_r}} c^{i}_{v_pv_r} 
\nonumber\\
&+& \sum_{v\neq v_1,v_p} \cB_v \Gamma_{\cB_v \cB^j_{vv'}} c^j_{vv'}
 \;.
\eea
Using $\Gamma_{\cB_{v_p}\cB^i_{v_1v_p}}=-\Gamma_{\cB_{v_1}\cB^i_{v_1v_p}}=\pm 1$, we rewrite equation (\ref{eq:terms}) as
\bea\label{eq:terms1}
d_1\alpha_1&=& (\cB_{v_1}-\cB_{v_p}) \Gamma_{\cB_{v_1}\cB^i_{v_1v_p}} c^{i}_{v_1v_p}+
\sum_{v_q\neq v_p} (\cB_{v_1} -\cB_{v_p})\Gamma_{\cB_{v_1}\cB^i_{v_1v_q}} c^{i}_{v_1v_q} \nonumber\\
&+& 
\sum_{v_q\neq v_p} \cB_{v_p} \Gamma_{\cB_{v_1}\cB^i_{v_1v_q}} c^{i}_{v_1v_q} +
\sum_{v_r\neq v_1} \cB_{v_p} \Gamma_{\cB_{v_p}\cB^i_{v_pv_r}} c^{i}_{v_pv_r} 
\nonumber\\
&+& \sum_{v\neq v_1,v_p} \cB_v \Gamma_{\cB_v \cB^j_{vv'}} c^j_{vv'}
 \;,
\eea
and perform the change of basis in $C_0(\cG)$
\bea
 \cB_{v_1}'=\cB_{v_1}-\cB_{v_p} ,\quad \cB_{v_q}'= \cB_{v_q},\; v_q\neq v_1 \; ,
\eea
under which equation (\ref{eq:terms}) becomes
\bea\label{eq:changevar}
d_1\alpha_1&=& \cB_{v_1}'\Gamma_{\cB_{v_1}\cB^i_{v_1v_p}} c^{i}_{v_1v_p}+ \sum_{v_q\neq v_p} \cB_{v_1}' \Gamma_{\cB_{v_1}\cB^i_{v_1v_q}} c^{i}_{v_1v_q} \nonumber\\
&+& \sum_{v_q\neq v_p} \cB_{v_p}' \Gamma_{\cB_{v_1}\cB^i_{v_1v_q}} c^{i}_{v_1v_q} 
+\sum_{v_r\neq v_1} \cB_{v_p}' \Gamma_{\cB_{v_p}\cB^i_{v_pv_r}} c^{i}_{v_pv_r} 
\nonumber\\
&+& \sum_{v\neq v_1,v_p} \cB_v \Gamma_{\cB_v \cB^j_{vv'}} c^j_{vv'}
\; .
\eea

The first line of equation (\ref{eq:changevar}) is the only one involving $\cB_{v_1}'$, which is linearly independent from all other $\cB_v'$, therefore it spans a direction in $\text{Im}(d_1)$. As $c^{i}_{v_1v_p}\in \ZZ$ and $\Gamma_{\cB_{v_p}\cB^i_{v_1v_p}}=\pm 1$, we have
\bea
 \text{Im}(d_1)=\mathbb{Z}\oplus \dots
\eea

The second and third lines of eq. (\ref{eq:changevar}) corespond to the incidence matrix of a graph in which all vertices $\cB_{v_q}$ who were connected by lines to $\cB_{v_1}$ are now connected by lines with $\cB_{v_p}$, that is the graph obtained from $\cG$ by contracting the line $\cB^i_{v_1v_p}$ and gluing the two vertices $\cB_{v_1}$ and $\cB_{v_p}$ into an unique vertex.

We iterate this contraction procedure for a spanning tree, that is $N-1$ times. Once such a tree is contracted, the final graph has only one vertex (rosette) $v$ and all remaining lines are loop lines. For any remaining loop line, the coefficient in this final sum of $c^j_{vv'}$ will be $\Gamma_{\cB_{v} \cB^j_{vv'}}+ \Gamma_{\cB_{v'} \cB^j_{vv'}} =0$. 

Therefore the image of $d_1$ is precisely
\bea
 \text{Im}(d_1)=\bigoplus_{N-1}\ZZ \Rightarrow  H_0=\frac{\text{ker}(d_0)}{\text{Im} (d_1)} =\ZZ
\; ,
\eea
and the kernel of $d_1$ is  
\bea \label{eq:kerd1}
\text{ker}(d_1)=\bigoplus_{L-(N-1)}\ZZ \; .
\eea

\qed

For the maximal homology group of a graph $\cG$ a similar result holds. In fact
a contraction procedure for a spanning tree in the graph implied that the first homotopy group is $H_0=\ZZ$, and a contraction procedure for a spanning tree in the dual space-time of a graph will in turn imply that $H_n=\ZZ$. 

Note first that $\text{Im}(d_{n+1})=0$. We have the lemma

\begin{lemma} \label{lem:max}
For a connected closed $n$D graph $\cG$  
 \bea
 \text{ker}(d_n)=\ZZ,\quad \text{Im}(d_n)=\bigoplus_{|\mB^n|-1}\ZZ \; .
 \eea
\end{lemma}

{\bf Proof:}
Denote the set of all colors $\mathfrak{C}=0,1,\dots, n$ and consider the boundary of an arbitrary $n$ chain
\bea \label{eq:maxchain}
&&d_n\Big{[} \sum_{\cB^{\cC}_{\cV}\in \mB^n} c^{\cC}_{\cV} \cB^{\cC}_{\cV}\Big{]}=
\sum_i \sum_{\stackrel{  \cB^{\cC}_{\cV}\in \mB^n}{\cC=\mathfrak{C}\setminus i}} 
c^{\cC}_{\cV} d_n\cB^{\cC}_{\cV}
\\
&& =\sum_i \sum_{\stackrel{  \cB^{\cC}_{\cV}\in \mB^n}{\cC=\mathfrak{C}\setminus i}} 
c^{\cC}_{\cV}
\sum_{j} \sum_{ \stackrel{\cB'^{\cC'}_{\cV'}\in \mB^{n-1}} {\cC'=\mathfrak{C}\setminus i\setminus j;
 \cV ' \subset \cV }} 
\Big{\{}(-)^{j+1} \cB'^{\cC'}_{\cV'}|_{j<i}+ (-)^{j} \cB'^{\cC'}_{\cV'}|_{j>i} \Big{\}}\nonumber\\
&&= \sum_{i,j; j<i}\sum_{\stackrel{\cB'^{\cC'}_{\cV'}\in \mB^{n-1}}{\cC'=\mathfrak{C}\setminus i\setminus j}}
\cB'^{\cC'}_{\cV'} \Big{[}(-)^{j+1}
c^{\cC}_{\cV}|_{\stackrel{\cV \supset \cV'}{\cC=\mathfrak{C}\setminus i}} 
+(-)^{i}
c^{\cC}_{\cV}|_{\stackrel{\cV \supset \cV'}{\cC=\mathfrak{C}\setminus j}} 
\Big{]} \; . \nonumber
\eea
Renaming the $c^{\cC}_{\cV}=(-)^i c^{\cC}_{\cV}$ if $\cC=\mathfrak{C}\setminus i$ and eq. (\ref{eq:maxchain}) becomes
\bea \label{eq:ready}
\sum_{i,j; j<i}\sum_{\stackrel{\cB'^{\cC'}_{\cV'}\in \mB^{n-1}}{\cC'=\mathfrak{C}\setminus i\setminus j}}
\cB'^{\cC'}_{\cV'} (-)^{i+j+1} \Big{[}
c^{\cC}_{\cV}|_{\stackrel{\cV \supset \cV'}{\cC=\mathfrak{C}\setminus i}} 
-c^{\cC}_{\cV}|_{\stackrel{\cV \supset \cV'}{\cC=\mathfrak{C}\setminus j}} 
\Big{]} \;.
\eea

First, note that the eq. (\ref{eq:ready}) is zero if and only if all $c^{\cC}_{\cV}$ are equal, hence
\bea
 \text{ker}(d_n)=\ZZ \Rightarrow H_n= \ZZ \; .
\eea

To determin the image of $d_n$, consider the space-time $\tilde{\cal G}$ dual to the GFT graph $\cG$. It is formed by vertices dual to the $n$D bubbles of $\cG$ and lines dual to the $n-1$D bubbles. We orient the space-time line dual to the bubble $\cB'^{\cC'}_{\cV'}\in \mathfrak{B}^{n-1}$ with colors $\cC'=\mathfrak{C}\setminus i \setminus j$ and $j<i$ from the space-time vertex dual to $\cB^{\cC}_{\cV}\in \mathfrak{B}^{n}$ with $\cV' \subset \cV$ and colors $\cC=\mathfrak{C}\setminus i$ to the space time vertex dual to $\cB^{\cC}_{\cV}\in \mathfrak{B}^{n}$ with $\cV' \subset \cV$ and colors $\cC=\mathfrak{C}\setminus j $. The matrix of the operator $d_n$ is then the transposed of the incidence matrix of the space-time $\tilde{\cal G}$ dual to $\cG$
\bea
  \Lambda_{\cB'^{\cC'}_{\cB'} \cB^{\cC}_{\cV}}=
 \begin{cases}
 1, &\text{ if the line dual to } \cB'^{\cC'}_{\cV'} \text{ enters the vertex dual to }  
\cB^{\cC}_{\cV} \\
 -1, &\text{ if the line dual to } \cB'^{\cC'}_{\cV'} \text{ exits the vertex dual to }  
\cB^{\cC}_{\cV} \\
  0, &\text{else}
 \end{cases} \,,
\eea
and redefining $ \cB'^{\cC'}_{\cV'} =\cB'^{\cC'}_{\cV'} (-)^{i+j+1} $ eq. (\ref{eq:ready}) rewrites
\bea\label{eq:ready1}
 d_n\alpha_n=\sum_{\cB'^{\cC'}_{\cV'}}  \cB'^{\cC'}_{\cV'} \Lambda_{\cB'^{\cC'}_{\cV'} 
\cB^{\cC}_{\cV}}
 c^{\cC}_{\cV} \; .
\eea

In the dual space-time $\tilde{\cal G}$, the $c^{\cC}_{\cV}$'s are associated to vertices and $\cB'^{\cC'}_{\cV'}$ to lines. Therefore equation (\ref{eq:ready1}) has the same form as (\ref{eq:terms}), with the $c^{\cC}_{\cV}$ and $\cB'^{\cC'}_{\cV'}$ swapped!

Call $\tilde{\cal T}$ a rooted tree in the dual graph $\tilde{\cG}$, and for all dual vertices $\cB^{\cC}_{\cV}$ call $\tilde{\cB'}^{\tilde{\cC'}}_{\tilde{\cV'}}$ the dual tree line touching this dual vertex and going towards the root. Under the change of variables parallel to (\ref{eq:changevar}), performed this time on the $c^{\cC}_{\cV}$, the quadratic form (\ref{eq:ready1}) becomes
\bea
d_n\alpha_n=\sum_{\cB^{\cC}_{\cV}\in \mathfrak{B}^{n}}c^{\cC}_{\cV}
\Big{[}\Lambda_{\tilde{\cB'}^{\tilde{\cC}'}_{\tilde{\cV}'} \cB^{\cC}_{\cV}} 
\tilde{\cB'}^{\tilde{\cC}'}_{\tilde{\cV}'} +\sum_{\cB'^{\cC'}_{\cV'}\notin \tilde{T}} 
 \Lambda_{\cB'^{\cC'}_{\cB'} \cB^{\cC}_{\cV}} \cB'^{\cC'}_{\cB'}
\Big{]} \; .
\eea
The vectors in brackets are linearly independent as each of them contains a  
$\tilde{\cB'}^{\tilde{\cC}'}_{\tilde{\cV}'}$ corresponding to a different tree line in 
$\tilde{\cal T}$. As 
$ \Lambda_{\tilde{\cB'}^{\tilde{\cC}'}_{\tilde{\cV}'} \cB^{\cC}_{\cV}}=\pm 1$,
$|\tilde{\cal T}|=|\mB^n|-1$ and $c^{\cC}_{\cV}\in \ZZ $ we conclude that 
\bea \label{eq:imdn}
 \text{Im}(d_n)=\bigoplus_{|\mB^n|-1}\ZZ \; .
\eea

\qed

The other homology groups $H_p$, $0<p<n$ depend of the particular colored GFT nD graph one analyzes. Note that in 3D, in order to compute $H_1$ and $H_2$ one needs only to determine the kernel and image of the operator $d_2$. Several examples are presented in the appendix (\ref{sec:app}).

\section{Amplitudes and Homotopy}\label{sec:homotopy}

After the study of the homology groups of a GFT graph, the next natural step is to study its homotopy groups. We will only define here a homotopy equivalence for curves which will enable us to define the fundamental group of a GFT graph.

In strict parallel to triangulated polyhedra, we define an edge path as an ordered sequence of vertices $[v_n,\dots v_1]$ connected by lines (that is, $\forall i$, $\exists \cB^{l}_{v_{i+1}v_i}$). An edge loop is a closed edge path, $v^n=v^1$.The paths 
\bea
 [v^{i+k},v^{i+k-1}, \dots ,v^{i+1}, v^i] \text{ and }
 [v^{i+k} v^{i+k+1}, \dots v^n, v^1,\dots v^{i-1}, v^i] \;,
\eea
are homotopically equivalent if the set of vertices $v^n,\dots v^1$ span a 2 bubble.

In strict parallel to triangulated polyhedra, we construct the edge path group of a colored GFT graph as follows. Start by associating group elements $g\in G$ to all lines of the graph. For all faces $\cB^{ab}_{\cV}$ of the graph, the set of vertices $\cV=v_n,\dots v_1$ is an closed path.  The {\it relations} defining the fundamental group are associated to the face of the graph and write 
\bea \label{eq:relations}
{\cal R}_{\cB^{ab}_{\cV}}= \prod_{\cB^{j}_{v_{i+1}v_{i}}\in \cB^{ab}_{\cV}} g_{\cB^{j}_{v_{i+1}v_{i}}}^{\sigma(\cB^{i}_{v_{i+1}v_{i}})}=e \;,
\eea
where $e$ is the unit element of $G$, and $\sigma(\cB^{i}_{v_{i+1}v_i})$ is 1 (or -1) if the vertex $v_i$ is positive (or negative). Finally, we set the group elements associated to a spanning tree ${\cal T}$ in the graph to $e$. The fundamental group of the graph is the group of words with generators $g_{\cB^{j}_{v_{i+1}v_{i} }}, \cB^{j}_{v_{i+1}v_{i}} \notin {\cal T}$ and relations (\ref{eq:relations}).

On the other hand, as proved in \cite{FreiGurOriti} for a more general case, the amplitude of a colored GFT graph is
\bea
 {\cal A}_{\cG}=\int [dg] \prod_{\cB^{ab}_{\cV}} \delta({\cal R}_{\cB^{ab}_{\cV}})
=\int_{g \notin {\cal T}} [dg] \prod_{\cB^{ab}_{\cV}} \delta({\cal R}_{\cB^{ab}_{\cV}}) \; ,
\eea
for any tree ${\cal T} \in \cG$. Therefore the Feynman amplitude of a graph is the volume of the relations defining the fundamental group of the graph over the base GFT group $G$! In this light the main result of \cite{FreiGurOriti} can be translated as follows (see \cite{FreiGurOriti} for the appropriate definitions): ``type 1 graphs dual to manifold space-times are homotopically trivial''. It is however difficult to give a complete characterization of homotopically trivial graphs. A first step in this direction is given by the lemma \ref{lem:genus} below.

\begin{lemma}\label{lem:genus}
 If a closed 3D graph is homotopically trivial, then all its 3 colors bubbles are planar.
\end{lemma}

{\bf Proof:} If a graph is homotopically trivial then it is homologically trivial $H_1=0$. Eq. (\ref{eq:kerd1}) then implies
\bea
 \text{Im}(d_2)=\bigoplus_{L-(N-1)}\ZZ \; .
\eea
As $d_2$ is defined on $\mathfrak{B}^2$, with $|\mathfrak{B}^2|=F$, and $\text{ker}(d_2)\supset \text{Im}(d_3)$ we conclude that
\bea\label{eq:homology}
 F-[L-(N+1)]\ge B-1 \; ,
\eea
where $B=|\mathfrak{B}^3|$.

On the other hand we will show below that for a 3D graph
\bea\label{eq:genus}
 N-L+F-B=-\sum_{\cB^{\cC}_{\cV}\in \mathfrak{B}^3} g_{\cB^{\cC}_{\cV}} \; ,
\eea
where $ g_{\cB^{\cC}_{\cV}}$ is the genus of the 3 colors bubble $\cB^{\cC}_{\cV}$. Eq. (\ref{eq:genus}) and (\ref{eq:homology}) imply that $g_{\cB^{\cC}_{\cV}}=0, \forall \cB^{\cC}_{\cV}\in \mathfrak{B}^3 $, therefore all the 3D bubbles are planar.

To prove (\ref{eq:genus}), consider all the 3D bubbles of a graph, that is all the connected components formed by lines of three fixed colors. Any vertex $v$ of the initial graph will appear four times in the four distinct bubbles $\cB^{012}_{\cV}$, $\cB^{013}_{\cV'}$, $\cB^{023}_{\cV''}$, $\cB^{123}_{\cV'''}$, such that $v\in \cV$, $v\in \cV'$, $v\in \cV''$, $v\in \cV'''$. Any line, say $\cB^{0}_{v_1v_2}$, will appear only three times, in the three bubbles $\cB^{012}_{\cV}$, $\cB^{013}_{\cV'}$, $\cB^{023}_{\cV''}$, but not in $\cB^{123}_{\cV'''}$! Faces have two colors hence appear only twice, in two bubbles. Denoting $n,l,f$ the total number of vertices lines and faces of all the bubbles, and $N,L,F$ the number of vertices lines and faces of the initial graph, we then have
\bea \label{eq:3D2D}
 n=4N,\quad l=3L,\quad f=2F \; .
\eea

For any 3D bubble $\cB^{\cC}_{\cV}$
\bea\label{eq:genuss}
 n_{\cB^{\cC}_{\cV}}-l_{\cB^{\cC}_{\cV}}+f_{\cB^{\cC}_{\cV}}=2-2g_{\cB^{\cC}_{\cV}} \, ,
\eea
and adding the equations (\ref{eq:genuss}) for all bubbles and using eq. (\ref{eq:3D2D}) gives
\bea \label{eq:almo}
&& n-l+f=2 |\cB^3|-2\sum_{\cB^{\cC}_{\cV}\in \mathfrak{B}^3} g_{\cB^{\cC}_{\cV}}\nonumber\\
&&\rightarrow 4N-3L+2F-2B=-2\sum_{\cB^{\cC}_{\cV}\in \mathfrak{B}^3} g_{\cB^{\cC}_{\cV}}\;.
\eea
As the vertices of the 3D graphs are four valent we have $2N=L$ which together with (\ref{eq:almo}) yields (\ref{eq:genus}).

\qed

This lemma implies not only that a homotopically trivial graph is dual to a manifold space-time, but also that the combinatorial complex of a homotopically trivial graph is naturally a CW complex.

Note that the reciprocal of lemma \ref{lem:genus} is not true. The first example in the appendix \ref{sec:app} is in fact a non homotopically trivial graph (actually not even homologically trivial!) whose bubbles are all planar.

\section{Conclusion}\label{sec:conclusion}

In this paper we introduced a new, fermionic group field theory. Using and appropriate boundary operator we proved that its graphs are combinatorial complexes. We defined and analyzed the associated homology. Furthermore we defined the natural homotopy transformation for paths on graphs and related the Feynman amplitude of graphs with the fundamental group.

A large amount of work should now be carried out in several directions. In a more quantum field theoretical approach one should not only continue the preliminary studies on the power counting of GFT's \cite{FreiGurOriti,Magnen:2009at} but also look for nontrivial fermionic instanton solutions corresponding to  \cite{Fairbairn:2007sv} and there possible relation with matter fields.

On the other hand our results could be used as a purely mathematical tool to further the understanding of three dimensional topological spaces. The encoding of the bubble complex in colored graphs provides a bridge between topological and combinatorial notions opening up the possibility to obtain, using the latter, new results on the former.

\section*{Acknowledgements}

The author would like to express his deepest thanks to Matteo Smerlak. Not only he pointed out to us the body of literature on manifold crystallization but also our numerous discussions on the topology of GFT graphs have been hugely beneficial for this work.

Research at Perimeter Institute is supported by the Government of Canada through Industry 
Canada and by the Province of Ontario through the Ministy of Research and Innovation.

\appendix

\section{Homology Computations} \label{sec:app}

In this appendix we compute the homology groups for three examples of four colored graphs.

\subsection{First Example}
\begin{figure}[htb]
\centering{
\includegraphics[width=40mm]{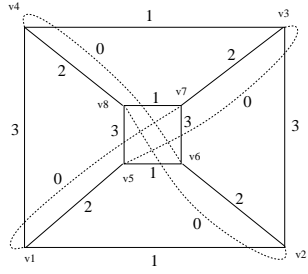}}
\caption{A first example of 3D graph.}
\label{fig:graph1}
\end{figure}

Consider the four colored graph of figure \ref{fig:graph1}. The reader can check that all lines connect a positive and a negative vertex. 

The 3D bubbles of this graph are
$\cB^{012}_{v_1\dots v_8}$, 
$\cB^{013}_{v_1\dots v_8}$, 
$\cB^{023}_{v_1\dots v_8}$ and  
$\cB^{123}_{v_1\dots v_8}$.
The 2D bubbles (faces) are
\bea
&&\cB^{01}_{v_1v_2v_8v_7}, \quad
\cB^{01}_{v_3v_4v_6v_5}, \quad 
\cB^{02}_{v_1v_7v_3v_5}, \quad
\cB^{02}_{v_4v_6v_2v_8}, \nonumber\\
&&\cB^{03}_{v_1v_7v_6v_4}, \quad
\cB^{03}_{v_2v_8v_5v_3}, \quad
\cB^{12}_{v_1v_2v_6v_5}, \quad
\cB^{12}_{v_3v_4v_8v_7} ,\nonumber\\
&&\cB^{13}_{v_1v_2v_3v_4}, \quad
\cB^{13}_{v_5v_6v_7v_8}, \quad
\cB^{23}_{v_1v_5v_8v_4}, \quad
\cB^{23}_{v_2v_3v_7v_6} .
\eea
The 1D bubbles are
\bea
&&\cB^{0}_{v_1v_7},\;
\cB^{0}_{v_2v_8}, \;
\cB^{0}_{v_3v_5}, \;
\cB^{0}_{v_4v_6}, \quad
\cB^{1}_{v_1v_2}, \;
\cB^{1}_{v_3v_4}, \;
\cB^{1}_{v_5v_6}, \;
\cB^{1}_{v_7v_8}, \nonumber\\
&&\cB^{2}_{v_1v_5}, \;
\cB^{2}_{v_2v_6}, \;
\cB^{2}_{v_3v_7}, \;
\cB^{2}_{v_4v_8}, \quad 
\cB^{3}_{v_1v_4}, \;
\cB^{3}_{v_2v_3}, \;
\cB^{3}_{v_5v_8}, \;
\cB^{3}_{v_6v_7} . 
\eea
while the 0D bubbles are vertices. We need to analyze the kernel and the image of the operator $d_2$. Acting on a two chain $d_2$ writes
\bea
d_2\alpha_2&=&c^{01}_{1287}[\cB_{v_1v_2}^1+ \cB_{v_7v_8}^1 - \cB_{v_1v_7}^0-
\cB_{v_2v_8}^0]\nonumber\\
&+&c^{01}_{3465}[\cB_{v_3v_4}^1+\cB_{v_5v_6}^1 - \cB_{v_3v_5}^0-
\cB_{v_4v_6}^0] 
\nonumber\\
&+&c^{02}_{1735}[\cB_{ v_1v_5}^2+ \cB_{v_3v_7}^2 - \cB_{v_1v_7}^0-
\cB_{ v_3v_5}^0]\nonumber\\
&+&c^{02}_{4628}[\cB_{v_2v_6}^2+ \cB_{ v_4v_8}^2 - \cB_{v_4v_6}^0-
\cB_{v_2v_8}^0] \nonumber\\
&+&c^{03}_{1764}[\cB_{ v_1v_4}^3+ \cB_{ v_6v_7}^3 - \cB_{ v_1v_7}^0-
\cB_{ v_4v_6}^0]\nonumber\\
&+&c^{03}_{2853}[\cB_{v_2v_3}^3+ \cB_{ v_5v_8}^3 - \cB_{v_2v_8}^0-
\cB_{ v_3v_5}^0] \nonumber\\
&+&c^{12}_{1265}[\cB_{ v_1v_5}^2+ \cB_{ v_2v_6}^2- \cB_{v_1v_2}^1
-\cB_{v_5v_6}^1]\nonumber\\
&+&c^{12}_{3487}[\cB_{v_3v_7}^2+ \cB_{ v_4v_8}^2 - \cB_{v_3v_4}^1
-\cB_{v_7v_8}^1] \nonumber\\
&+&c^{13}_{1234}[\cB_{v_1v_4}^3+ \cB_{ v_2v_3}^3 - \cB_{v_1v_2}^1-
\cB_{v_3v_4}^1]\nonumber\\
&+&c^{13}_{5678}[\cB_{v_5v_8}^3+ \cB_{ v_6v_7}^3 - \cB_{v_5v_6}^1-
\cB_{v_7v_8}^1] \nonumber\\
&+&c^{23}_{1584}[\cB_{v_5v_8}^3+ \cB_{ v_1v_4}^3 - \cB_{ v_1v_5}^2-
\cB_{v_4v_8}^2] \nonumber\\
&+&c^{23}_{2376}[\cB_{v_2v_3}^3+ \cB_{v_6v_7}^3 - \cB_{ v_2v_6}^2-
\cB_{ v_3v_7}^2] \; .
\eea
A lengthy but straightforward computation shows that
\bea
\text{Im}(d_2)=\bigoplus_8 \ZZ \oplus 2\ZZ,\quad \text{ker}d_2=\bigoplus_3 \ZZ \; .
\eea
Using lemma \ref{lem:d1} we conclude that $\text{ker}(d_1)=\bigoplus_{16-7}\ZZ$ and using lemma
\ref{lem:max} we conclude that $\text{Im}(d_3) = \bigoplus_{3-1} \ZZ$. Therefore
\bea
&& H_0=\ZZ \;, \nonumber\\
&& H_1=\frac{\text{ker} (d_1)}{\text{Im}(d_2)}
=\frac{\bigoplus_9 \ZZ}{\bigoplus_8 \ZZ \oplus 2\ZZ}=\ZZ_2 \;, \nonumber\\
&&H_2 =\frac{\text{ker} (d_2)}{\text{Im}(d_3)} = \frac{\bigoplus_3 \ZZ}{\bigoplus_3 \ZZ}=0 \;,
\nonumber\\
&&H_3 =\ZZ \; .
\eea

Note first that these homology groups match those of $\mathbb{R}P^3$. Second direct inspection shows that all the bubbles of this graph are planar.

\subsection{Second Example}

\begin{figure}[htb]
\centering{
\includegraphics[width=40mm]{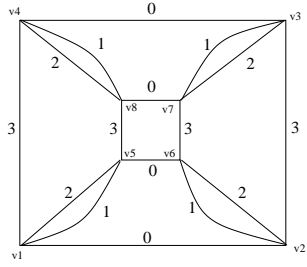}}
\caption{A second example of 3D graph.}
\label{fig:graph2}
\end{figure}

Consider now the graph of figure \ref{fig:graph2}.
 
The 3D bubbles of this graph are
$\cB^{012}_{v_1v_2v_6v_5}$, 
$\cB^{012}_{v_3v_4v_8v_7}$,
$\cB^{123}_{v_1v_5v_8v_4} $,
$\cB^{123}_{v_2v_3v_7v_6} $,
$ \cB^{013}_{v_1\dots v_8} $ and 
$\cB^{023}_{v_1\dots v_8}$.
The 2D bubbles are
\bea
&&\cB^{01}_{v_1v_2v_5v_6} , \;
\cB^{01}_{v_3v_4v_8v_7}, \;
\cB^{02}_{ v_1v_2v_5v_6}, \;
\cB^{02}_{v_3v_4v_8v_7}, \; 
\cB^{03}_{ v_1v_2v_3v_4}, \;
\cB^{03}_{ v_5v_6v_7v_8}, \\
&&\cB^{12}_{ v_1v_5}, \; \cB^{12}_{ v_2v_6}, \;
\cB^{12}_{ v_4v_8} ,\; \cB^{12}_{ v_3v_7},\;
\cB^{13}_{ v_1v_5v_8v_4} ,\; 
\cB^{13}_{ v_2v_3v_7v_6}, \;
\cB^{23}_{ v_1v_5v_8v_4} ,\; 
\cB^{23}_{ v_2v_3v_7v_6}. \nonumber
\eea

The 1D bubbles are
\bea
&&\cB^{0}_{v_1v_2}, \;
\cB^{0}_{v_5v_6}, \;
\cB^{0}_{v_3v_4}, \;
\cB^{0}_{v_7v_8},\;
\cB^{1}_{v_1v_5}, \;
\cB^{1}_{v_2v_6}, \;
\cB^{1}_{v_3v_7}, \;
\cB^{1}_{v_4v_8},  \nonumber\\
&&\cB^{2}_{v_1v_5}, \;
\cB^{2}_{v_2v_6}, \;
\cB^{2}_{v_3v_7}, \;
\cB^{2}_{v_4v_8}, \; 
\cB^{3}_{v_1v_4}, \;
\cB^{3}_{v_2v_3}, \;
\cB^{3}_{v_5v_8}, \;
\cB^{3}_{v_6v_7}. 
\eea
Lemmas \ref{lem:d1} and \ref{lem:max} give
\bea
 H_0= \ZZ,  \quad \text{ker}(d_1)=\bigoplus_9 \ZZ,\qquad 
 H_3=\ZZ, \quad \text{Im}(d_3)=\bigoplus_5\ZZ \; .
\eea

Manipulations similar to the ones for the first example give
\bea
\text{Im}(d_2)=\bigoplus_9 \ZZ \rightarrow H_1=0, \quad \text{ker}(d_2)=\bigoplus_5 \ZZ
\rightarrow H_2=0 \; .
\eea

These homology groups are consistent with those of $S^3$, and a direct computation of the Feynman amplitude shows that indeed this graph is homotopically trivial.

\subsection{Third Example}
\begin{figure}[htb]
\centering{
\includegraphics[width=40mm]{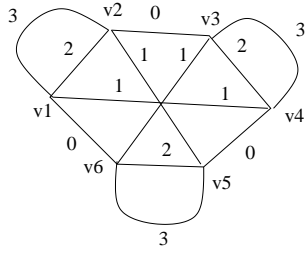}}
\caption{A third example of a graph.}
\label{fig:graph3}
\end{figure}

For the final example take the graph of figure \ref{fig:graph3}.

The 3D bubbles of this graph are
\bea
\cB^{012}_{v_1v_2v_3v_4v_5v_6}, \;
 \cB^{013}_{v_1v_2v_3v_4v_5v_6}, \;
 \cB^{023}_{v_1v_2v_3v_4v_5v_6}, \;
 \cB^{123}_{v_1v_2v_3v_4v_5v_6} \;.
\eea
Its 2D bubbles are
\bea 
&& \cB^{01}_{v_1v_2v_3v_4v_5v_6} ,\; \cB^{02}_{v_1v_2v_3v_4v_5v_6} ,\;
 \cB^{03}_{v_1v_2v_3v_4v_5v_6} ,\;
 \cB^{12}_{v_1v_2v_3v_4v_5v_6} ,\;
 \cB^{13}_{v_1v_2v_3v_4v_5v_6} , \nonumber\\
&& \cB^{23}_{v_1v_2}, \quad \cB^{23}_{v_3v_4}, \quad \cB^{23}_{ v_5v_6} \;,
\eea
and its 1D bubbles are
\bea
&& \cB^{0}_{v_1v_6}, \; \cB^{0}_{v_2v_3},  \; \cB^{0}_{v_4v_5}, \;
 \cB^{1}_{v_1v_4}, \; \cB^{1}_{v_2v_5},  \; \cB^{1}_{v_3v_6} , \nonumber\\
&& \cB^{2}_{v_1v_2}, \; \cB^{2}_{v_3v_4},  \; \cB^{2}_{v_5v_6}, \;
 \cB^{3}_{v_1v_2}, \; \cB^{3}_{v_3v_4},  \; \cB^{3}_{v_5v_6}\; . 
\eea
Again from \ref{lem:d1} and \ref{lem:max} we get
\bea
  H_0= \ZZ,  \quad \text{ker}(d_1)=\bigoplus_{7} \ZZ,\qquad 
 H_3=\ZZ, \quad \text{Im}(d_3)=\bigoplus_3\ZZ \; ,
\eea
and a direct computation gives
\bea
 \text{Im}(d_2)=\bigoplus_5 \ZZ \rightarrow H_1=\ZZ\oplus\ZZ ;
\quad \text{ker}(d_2)=\bigoplus_3\ZZ \rightarrow H_2=0 \; .
\eea

\end{document}